\documentclass[aps,prb,a4paper,superscriptaddress,twocolumn,floatfix,showpacs]{revtex4}
\usepackage{graphicx}
\usepackage{latexsym}
\usepackage[]{amsmath}

\begin{document}
\newcommand{\chem}[1]{\ensuremath{\mathrm{#1}}}

\title{Determination of characteristic muon precession and relaxation signals in 
\chem{FeAs} and \chem{FeAs_2}, possible impurity phases in pnictide superconductors}

\author{P.\ J.\ Baker}
\affiliation{Oxford University Department of Physics, Clarendon Laboratory,
Parks Road, Oxford OX1 3PU, United Kingdom}

\author{H.\ J.\ Lewtas}
\affiliation{Oxford University Department of Physics, Clarendon Laboratory,
Parks Road, Oxford OX1 3PU, United Kingdom}

\author{S.\ J.\ Blundell}
\affiliation{Oxford University Department of Physics, Clarendon Laboratory,
Parks Road, Oxford OX1 3PU, United Kingdom}

\author{T.\ Lancaster}
\affiliation{Oxford University Department of Physics, Clarendon Laboratory,
Parks Road, Oxford OX1 3PU, United Kingdom}

\author{F.\ L.\ Pratt}
\affiliation{ISIS Muon Facility, ISIS, Chilton, Oxon.,
OX11 0QX, United Kingdom}

\author{D.\ R.\ Parker}
\affiliation{Inorganic Chemistry Laboratory, University of Oxford,
South Parks Road, Oxford, OX1 3QR, United Kingdom}

\author{M.\ J.\ Pitcher}
\affiliation{Inorganic Chemistry Laboratory, University of Oxford,
South Parks Road, Oxford, OX1 3QR, United Kingdom}

\author{S.\ J.\ Clarke}
\affiliation{Inorganic Chemistry Laboratory, University of Oxford,
South Parks Road, Oxford, OX1 3QR, United Kingdom}

\date{\today}

\begin{abstract}
We report muon-spin relaxation measurements of highly homogeneous
samples of \chem{FeAs} and \chem{FeAs_2}, both previously found as
impurity phases in some samples of recently synthesized pnictide
superconductors.  We observe well defined muon precession in the
\chem{FeAs} sample with two precession frequencies of $38.2(3)$ and
$22.7(9)$~MHz at $7.5$~K, with the majority of the amplitude
corresponding to the lower frequency component.  In \chem{FeAs_2} we
confirm previous measurements showing that no long-ranged magnetic
order occurs above $2$~K and measure the muon spin relaxation rate,
which increases on cooling. Our results exclude the possibility that
previous muon-spin relaxation measurements of pnictide superconductors
have been measuring the effect of these possible impurities.
\end{abstract}


\maketitle

With the recent discovery of high-temperature superconductivity in
pnictide compounds containing \chem{FeAs} layers~\cite{kamihara08}
considerable research activity has been devoted to studying both their
superconducting and magnetic properties.  A rich variety of compounds
related to the original \chem{LaFeAsO_{\rm 1-x}F_{\rm x}} have already
been synthesized: 1111
compounds~\cite{chen08nature,chen08prl,ren08epl} where the
\chem{La^{3+}} ion has been substituted for other rare-earths leading
to a higher superconducting transition temperature, $T_c \sim 55$~K;
122 compounds~\cite{rotter08,sasmal08arxiv} related to the parent
compound \chem{BaFe_{2}As_{2}}
(e.g. \chem{Ba_{0.6}K_{0.4}Fe_{2}As_{2}}: $T_c = 38$~K), and
\chem{LiFeAs} where critical temperatures between 12 K and 18 K have
been reported.~\cite{wang08arxiv,pitcher08arxiv,tapp08prb} Most show
some competition between superconductivity and magnetism as variables
such as pressure or doping are changed. As these new materials have
been characterized it has become apparent that some samples contain
significant concentrations of magnetic
impurities~\cite{nowik08arxiv,sidorenko08arxiv} which could affect
their measured properties. Previously determined M\"{o}ssbauer
results~\cite{yuzuri80} on the three most plausible pnictide impurity
phases: \chem{FeAs}, \chem{FeAs_2}, and \chem{Fe_{2}As} have been
compared to spectra from a series of \chem{FeAs}-based
superconductors, suggesting that the impurity concentration could be
up to $50$\% in some samples.~\cite{nowik08arxiv} The $^{75}$\chem{As}
NMR results on \chem{SmFeAsO_{\rm 1-x}F_{\rm x}} showed a resonance
line at $265$~MHz that was attributed to the presence of the binary
phase \chem{FeAs}, possibly on a nanoscopic
scale.~\cite{sidorenko08arxiv} It is therefore important that the
properties of possible impurity phases are known so that the quality
of existing and future results can be accurately assessed.

Among the techniques already applied to the pnictide superconductors
has been muon-spin relaxation ($\mu$SR),~\cite{blundell99,sonier00}
measuring both the superconducting and magnetic properties.
\cite{luetkens08prl,drew08prl,carlo08arxiv,luetkens08arxiv,khasanov08arxiv,takeshita08arxiv,drew08arxiv,aczel08arxiv,goko08arxiv}
Two sorts of magnetic effects have been reported: static \chem{Fe}
magnetism, usually in the parent compound or lightly-doped samples;
and fluctuating magnetism in more strongly doped samples. The presence
of lanthanide moments can also play a r\^{o}le at lower
temperatures.~\cite{drew08prl,khasanov08arxiv}

Static \chem{Fe} magnetism leads to muon precession below the magnetic
ordering temperature $T_{\rm N}$, which is typically $150$~K in the
parent compounds (i.e.  \chem{LaFeAsO} and \chem{BaFe_{2}As_{2}}) and
falls with doping. The precession frequencies reported for the 1111 and
122 compounds already measured are shown in
Table~\ref{tab:frequencies}.
In the absence of long range magnetic order,
fluctuating \chem{Fe} moments lead to muon spin relaxation without
coherent precession. Results on superconducting \chem{Sm}-containing
oxypnictides with~\cite{drew08prl} and without~\cite{khasanov08arxiv}
\chem{F}-doping have shown a significant thermally activated change in
the zero-field muon spin relaxation rate which has been argued to be
intrinsic to the samples. This has been disputed following NMR
measurements on notionally similar compositions in which \chem{FeAs}
impurity concentrations of up to 50\% were found.~\cite{nowik08arxiv}
Temperature dependent relaxations have also been found in a number of other 
oxypnictide samples.~\cite{takeshita08arxiv,carlo08arxiv}

\begin{table}[t]
\caption{\label{tab:frequencies}
Precession frequencies ($T\rightarrow 0$) reported in samples of 1111 and 122
\chem{FeAs}-based superconductors. For the compounds which show more than
one precession frequency the proportions of the signal amplitudes are given
in brackets.
}
\begin{ruledtabular}
\begin{tabular}{lcc}
Sample & Reference & Precession Frequencies (MHz) \\
\hline
1111 & \\
\hline
\chem{LaFeAsO} & Refs.~\onlinecite{klauss08prl,carlo08arxiv} & 23 (70-90\%), 3 (10-30\%) \\
\chem{LaFeAsO_{\rm 1-x}F_{\rm x}} & Ref.~\onlinecite{luetkens08prl} & $22-24$ \\
\chem{LaFeAsO_{0.97}F_{0.03}} & Ref.~\onlinecite{carlo08arxiv} & 18, 10, 2 \\
\chem{SmFeAsO} & Ref.~\onlinecite{drew08arxiv} & 23.8 \\
\chem{NdFeAsO} & Ref.~\onlinecite{aczel08arxiv} & $\sim 24$ \\
\chem{CeFeAsO_{\rm 0.94}F_{0.06}} & Ref.~\onlinecite{goko08arxiv} & 8 \\
\hline
122 & \\
\hline
\chem{BaFe_{2}As_{2}} & Ref.~\onlinecite{aczel08arxiv} & 28.8 (80\%), 7 (20\%) \\
\chem{Ba_{0.55}K_{0.45}Fe_{2}As_{2}} & Ref.~\onlinecite{aczel08arxiv} & 26 \\
\chem{Ba_{0.5}K_{0.5}Fe_{2}As_{2}} & Ref.~\onlinecite{goko08arxiv} & 18, 4 \\ 
\chem{SrFe_{2}As_{2}} & Ref.~\onlinecite{goko08arxiv} & 45 \\
\chem{Sr_{0.5}Na_{0.5}Fe_{2}As_{2}} & Ref.~\onlinecite{goko08arxiv} & 37, 10 \\ 
\end{tabular}
\end{ruledtabular}
\end{table}

Such measurements of the magnetic properties could indeed be strongly
affected by the presence of magnetic impurity phases and therefore it
is important to isolate possible impurity phases and measure their
properties in isolation.  The $\mu$SR technique allows the values of
magnetic fields at muon stopping sites to be determined very
precisely. Also, the amplitude of a signal resulting from a particular
phase is proportional to its volume fraction in the sample, since the
muon decay asymmetry is a sum over a large number of muons stopped in
different places within the sample. It is therefore possible to make a
direct comparison between the muon spin relaxation in pure samples of
the proposed impurity phases and the signal resulting from impurity
phases in an \chem{FeAs}-based superconducting or magnetic sample.  In
this Brief Report we study the form of the muon-spin relaxation in the
two principal impurity phases, \chem{FeAs} and
\chem{FeAs_2},~\cite{footnote1} in order to more clearly interpret
the results of previous reported $\mu$SR studies of magnetism in
\chem{FeAs}-based materials.

\begin{figure}[t]
\includegraphics[width=8.5cm]{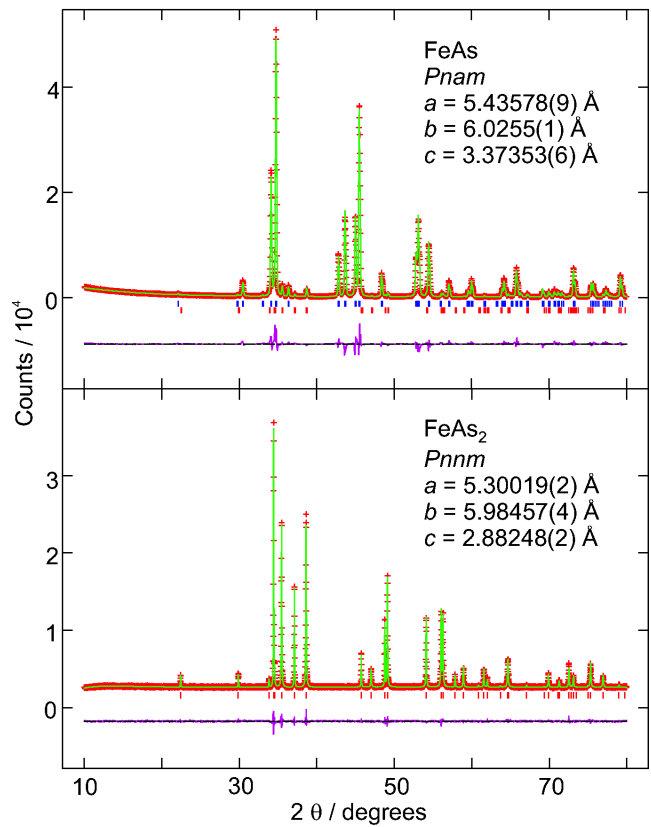}
\caption{ 
(Color online.) Rietveld refinements against laboratory PXRD
data for the FeAs (upper) and FeAs$_2$ (lower) samples used in the $\mu$SR
experiments. Data (red points), fit (green line) and difference
(purple line) are shown. In the case of FeAs tick marks are for 2 mol
\% FeAs$_2$ (lower) and the FeAs majority phase (upper). Space groups and
refined lattice parameters are included in the plots. Weighted profile
R-factors are 0.11 for the FeAs refinement (data collected in 18 hours
on a Rigaku Miniflex II diffractometer equipped with a secondary beam
monochromator using CuK$\alpha$ radiation) and 0.02 for the FeAs$_2$ refinement
(data collected in 3 hours on a PANalytical X-pert PRO diffractometer
using Cu K$\alpha_1$ radiation.)
\label{xray}}
\end{figure}

\chem{FeAs} is known to have a N\'{e}el temperature around 
$T_{\rm N} = 77$~K and displays helimagnetic order, as is common
for compounds with the \chem{MnP} structure.~\cite{selte72,lyman84}
\chem{FeAs_2} has previously been shown not to order above $5$~K by
M\"{o}ssbauer spectroscopy,~\cite{yuzuri80} being
diamagnetic.~\cite{fan72} FeAs and FeAs$_2$ were prepared by similar
methods. All manipulations were performed in an argon-filled
glovebox. Fe (Alfa Aesar, 99.998\%) and As (Alfa Aesar, 99.9999\%)
were mixed in the appropriate molar ratios, and sealed in evacuated
silica tubes. These vessels were heated for 48 hours at 500$^\circ$C,
then the samples were re-ground, resealed in fresh silica tubes and
annealed at 800$^\circ$C for a further 48 hours. The FeAs$_2$ sample
was found to be single phase by Powder X-ray Diffraction (PXRD),
although a small amount of elemental arsenic had sublimed away from
the solid in the reaction vessel. The FeAs sample was found to contain
1.9 mol \% FeAs$_2$ as an impurity phase. Rietveld refinements
(Fig.~\ref{xray}) against laboratory PXRD data produce lattice
parameters consistent with previous reports\cite{selte72,fan72} and do
not suggest any non-stoichiometry.
The samples are contaminated by very small amounts of
ferromagnetic impurity phases, but at levels below the sensitivity of
laboratory PXRD; this will not affect the $\mu$SR data (because muons
are a volume probe) but also mimics the trace impurities that are
likely to be present in pnictide samples which are prepared in similar ways.

Muon-spin rotation ($\mu$SR) experiments were performed on the General
Purpose Surface-Muon Instrument (GPS) at the Swiss Muon Source (Paul
Scherrer Institute, Switzerland).  In our $\mu$SR
experiments~\cite{blundell99} spin-polarized positive muons
(gyromagnetic ratio $\gamma_{\mu}/2\pi = 135.5$~MHzT$^{-1}$, lifetime
$\tau_{\mu} = 2.2~\mu$s) were implanted into the samples, where they
stop quickly without significant loss of spin polarization.  Each of
the powder samples was mounted inside a silver packet on a flypast
sample holder to reduce the background from muons stopping outside the
sample.  To measure the time evolution of the muon spin polarization,
emitted decay positrons were collected in detectors placed forward (F)
and backward (B) relative to the initial muon spin direction
(antiparallel to the beam momentum).  The muon decay asymmetry is
defined in terms of the count rates in the two detectors ($N_{\rm F}$
and $N_{\rm B}$) as:
\begin{equation}
A(t) = \frac{N_{\rm F}(t) - \alpha N_{\rm B}(t)}{N_{\rm F}(t) + \alpha N_{\rm B}(t)},
\label{asymmetry}
\end{equation}
where $\alpha$ is an experimental calibration constant related to the
relative efficiency of the detectors.

The muon spins are sensitive to both static and fluctuating local
fields at their stopping positions inside the material, and these
affect how the form of the muon decay asymmetry changes with time. In
\chem{FeAs} we see two distinct behaviors in different temperature
ranges and therefore the data require a different parameterization in
each temperature range. At low-temperature in \chem{FeAs} we have long
range magnetic order and quasistatic magnetic fields at the muon
stopping sites with two well separated precession frequencies. The
data are well described by the function:
\begin{eqnarray}
A(t) &=& \sum_{i=1,2} A_i e^{-\lambda_{i} t} \cos(2\pi\nu_i t) 
+ A_3 e^{-\lambda_3 t}.
\label{fafitfunc}
\end{eqnarray}
The two terms in the sum are damped oscillations corresponding to
muons precessing around quasistatic local fields at two magnetically
inequivalent stopping sites.  The $\lambda_i$ parameters describe 
damping rates and $\nu_i = \gamma_{\mu} B_i/2\pi$ are the
precession frequencies proportional to the magnitude of the magnetic
field at each stopping site. The third term is an exponential
relaxation, of rate $\lambda_3$, due to fluctuations flipping the spins
of muons having a non-zero spin component along the local magnetic field
direction.

For the paramagnetic phase of \chem{FeAs} 
the muon decay asymmetry was well described by a
single exponential relaxation:
\begin{equation}
A(t) = A_0 \exp(-\lambda t).
\label{faafitfunc}
\end{equation}

\begin{figure}[t]
\includegraphics[width=8.5cm]{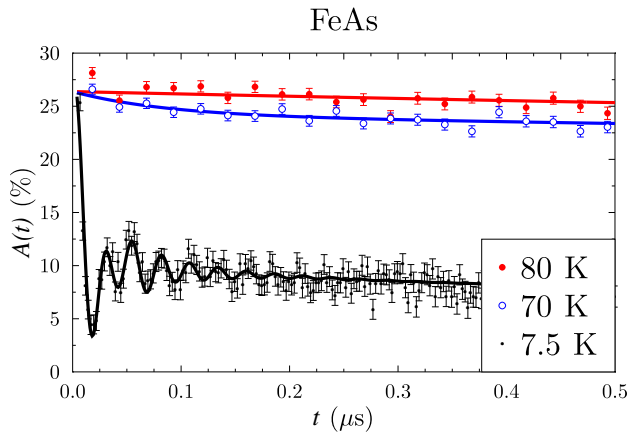}
\caption{
(Color online.) 
Muon asymmetry data for \chem{FeAs} showing the spin precession signal
evident at low-temperature, the paramagnetic signal at 80~K, and the rapid
spin depolarization evident close to $T_{\rm N}$. The low-temperature data
are fitted to Eq.~(\ref{fafitfunc}) and the $80$~K data are fitted to
Eq.~(\ref{faafitfunc}).
\label{FeAsrawdata}}
\end{figure}

The muon decay asymmetry data recorded for \chem{FeAs} at temperatures
of $7.5$, $70$, and $80$~K are shown in Fig.~\ref{FeAsrawdata}.  At
temperatures well below $T_{\rm N}$ we observe a well-defined muon
spin precession at two frequencies. At $7.5$~K the dominant frequency
is $\nu_1 = 22.7(9)$~MHz, accounting for around $70$\% of the
precession signal.  This frequency is strongly damped with $\lambda_1
= 52(6)$~MHz. The remaining $30$\% of the signal has a precession
frequency of $\nu_2 = 38.2(3)$~MHz with a damping rate of $\lambda_2 =
18(3)$~MHz. The exponentially relaxing part of the signal has a
relaxation rate of $\lambda_3 = 1.9(2)$~MHz.  This form of the
relaxation is evident up to close to $T_{\rm N}$ with similar
amplitudes as the temperature increases, but the damping of the
precession signals increases. Above $60$~K the precession is
overdamped and we can no longer resolve the precession frequencies.
We find no evidence of any additional components in the $\mu$SR
spectra below $T_{\mathrm{N}}$, leading us to believe that FeAs is
completely ordered throughout its bulk in this temperature regime.
Above $T_{\rm N}$ the relaxation rate decreases with increasing
temperature.

\begin{figure}[t]
\includegraphics[width=8.5cm]{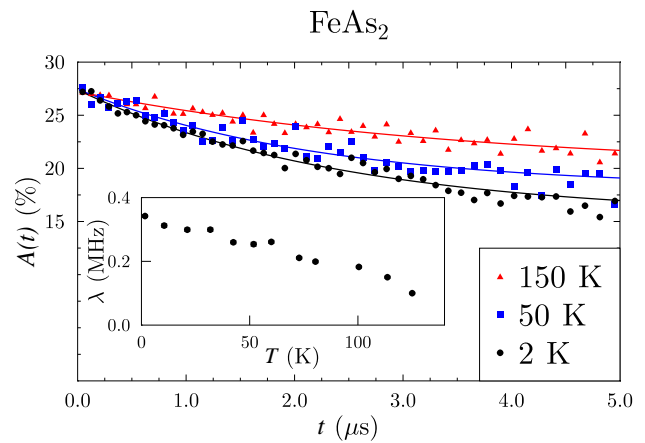}
\caption{
(Color online.) Muon asymmetry data for \chem{FeAs_2} showing the form of the
spin relaxation over the measured temperature range. The lines are fits to
Eq.~(\ref{faafitfunc}) described in the text. Inset: Temperature dependence
of the muon spin relaxation rate $\lambda$ defined in Eq.~(\ref{faafitfunc}).
\label{FeAs2}}
\end{figure}

Examples of the muon decay asymmetry data recorded on the
\chem{FeAs_2} sample are shown in Fig.~\ref{FeAs2}. The data are well
described by a single exponential relaxation [Eq.~(\ref{faafitfunc})]
at all temperatures, but the relaxation rate increases on cooling to
around $0.35$~MHz at $2$~K.

We can now compare our results to those summarized in the introduction
and Table~\ref{tab:frequencies}. None of the samples previously
measured using $\mu$SR show both the combination of frequencies we
have measured or the relative amplitudes we have measured for
\chem{FeAs}. The $23$~MHz signal we observe is similar to that in the
1111 compounds but they have no higher frequency signal, and a much
lower damping rate. The $23$~MHz signal in the 1111 compounds also
persists all the way up to $T_{\rm N} \sim 150$~K, twice that of
\chem{FeAs}. From this we must conclude that the local environments
for implanted muons are relatively similar in \chem{FeAs} and the 1111
compounds, but \chem{FeAs} impurities are not producing this
signal. The precession frequencies are significantly different from
those in the 122 materials and they show magnetic transitions at $T >
140$~K.  The $3$~MHz signal reported~\cite{klauss08prl,carlo08arxiv}
in \chem{LaFeAsO} is not due to \chem{FeAs}, despite the onset around
$70$~K reported in Ref.~\onlinecite{klauss08prl}.  In \chem{Sm}-based
oxypnictides a temperature dependent relaxation behavior was
observed~\cite{drew08prl,khasanov08arxiv} which followed a thermally
activated behavior consistent with crystal field excitations to the
first excited state of the \chem{Sm^{3+}} ion. On the basis of
M\"{o}ssbauer measurements on nominally similar samples this has been
argued to be due to \chem{FeAs} impurities.~\cite{nowik08arxiv} Our
measurements on \chem{FeAs} strongly exclude this possibility. Since
the measurements on \chem{F}-doped samples were performed at a pulsed
muon source at which the pulse length places an upper limit on the
measureable precession frequency any \chem{FeAs} impurities would have
given a drop in the initial muon asymmetry, $A(0)$, of $~0.15$\% per
$1$\% of impurity present.~\cite{footnote2} No such drop in the
initial asymmetry was observed.~\cite{drew08prl} The experiment of
Ref.~\onlinecite{khasanov08arxiv} would have measured signals very
similar to those in Fig.~\ref{FeAsrawdata} had any \chem{FeAs} been
present in the samples, which they did not. From this comparison we
conclude that none of the \chem{FeAs}-based superconductors previously
measured using $\mu$SR were sufficiently contaminated with \chem{FeAs} 
inclusions so as to give results directly due to such an impurity.

The situation in those 1111 compounds where a temperature dependent
relaxation rate is reported is more complicated to relate to impurity
phases. None of the samples show the same temperature dependent
relaxation rate as \chem{FeAs_2}, although there is relatively little
data taken around $100$~K. \chem{FeAs_2} certainly does not show the
same thermally activated temperature dependence observed in the
\chem{Sm}-oxypnictides,~\cite{drew08prl,khasanov08arxiv} nor is it
showing any significant changes in the relaxation rate in the lower
temperature region where features have been found in other
oxypnictides.~\cite{carlo08arxiv,takeshita08arxiv} That the values of
the relaxation rates in the impurity phases are similar to those in
the oxypnictides is more likely to be related to the similarity of the
muon's local environment in the superconductors to that in the
magnetic systems we consider here.

Part of this work was performed at the Swiss Muon Source, Paul
Scherrer Institute, Villigen, CH. We are grateful to Hubertus Luetkens
for experimental assistance and to the EPSRC (UK) for financial
support.

\bibliography{pjb-feas}

\begin{thebibliography}{29}
\expandafter\ifx\csname natexlab\endcsname\relax\def\natexlab#1{#1}\fi
\expandafter\ifx\csname bibnamefont\endcsname\relax
  \def\bibnamefont#1{#1}\fi
\expandafter\ifx\csname bibfnamefont\endcsname\relax
  \def\bibfnamefont#1{#1}\fi
\expandafter\ifx\csname citenamefont\endcsname\relax
  \def\citenamefont#1{#1}\fi
\expandafter\ifx\csname url\endcsname\relax
  \def\url#1{\texttt{#1}}\fi
\expandafter\ifx\csname urlprefix\endcsname\relax\def\urlprefix{URL }\fi
\providecommand{\bibinfo}[2]{#2}
\providecommand{\eprint}[2][]{\url{#2}}

\bibitem[{\citenamefont{Kamihara et~al.}(2008)\citenamefont{Kamihara, Watanabe,
  Hirano, and Hosono}}]{kamihara08}
\bibinfo{author}{\bibfnamefont{Y.}~\bibnamefont{Kamihara}},
  \bibinfo{author}{\bibfnamefont{T.}~\bibnamefont{Watanabe}},
  \bibinfo{author}{\bibfnamefont{M.}~\bibnamefont{Hirano}}, \bibnamefont{and}
  \bibinfo{author}{\bibfnamefont{H.}~\bibnamefont{Hosono}},
  \bibinfo{journal}{J.\ Am.\ Chem.\ Soc.} \textbf{\bibinfo{volume}{130}},
  \bibinfo{pages}{3296} (\bibinfo{year}{2008}).

\bibitem[{\citenamefont{Chen et~al.}(2008{\natexlab{a}})\citenamefont{Chen, Wu,
  Wu, Liu, Chen, and Fang}}]{chen08nature}
\bibinfo{author}{\bibfnamefont{X.~H.} \bibnamefont{Chen}},
  \bibinfo{author}{\bibfnamefont{T.}~\bibnamefont{Wu}},
  \bibinfo{author}{\bibfnamefont{G.}~\bibnamefont{Wu}},
  \bibinfo{author}{\bibfnamefont{R.~H.} \bibnamefont{Liu}},
  \bibinfo{author}{\bibfnamefont{H.}~\bibnamefont{Chen}}, \bibnamefont{and}
  \bibinfo{author}{\bibfnamefont{D.~F.} \bibnamefont{Fang}},
  \bibinfo{journal}{Nature} \textbf{\bibinfo{volume}{453}},
  \bibinfo{pages}{761} (\bibinfo{year}{2008}{\natexlab{a}}).

\bibitem[{\citenamefont{Chen et~al.}(2008{\natexlab{b}})\citenamefont{Chen, Li,
  Wu, Li, Hu, Dong, Zheng, Luo, and Wang}}]{chen08prl}
\bibinfo{author}{\bibfnamefont{G.~F.} \bibnamefont{Chen}},
  \bibinfo{author}{\bibfnamefont{Z.}~\bibnamefont{Li}},
  \bibinfo{author}{\bibfnamefont{D.}~\bibnamefont{Wu}},
  \bibinfo{author}{\bibfnamefont{G.}~\bibnamefont{Li}},
  \bibinfo{author}{\bibfnamefont{W.~Z.} \bibnamefont{Hu}},
  \bibinfo{author}{\bibfnamefont{J.}~\bibnamefont{Dong}},
  \bibinfo{author}{\bibfnamefont{P.}~\bibnamefont{Zheng}},
  \bibinfo{author}{\bibfnamefont{J.~L.} \bibnamefont{Luo}}, \bibnamefont{and}
  \bibinfo{author}{\bibfnamefont{N.~L.} \bibnamefont{Wang}},
  \bibinfo{journal}{Phys.\ Rev.\ Lett.} \textbf{\bibinfo{volume}{100}},
  \bibinfo{pages}{247002} (\bibinfo{year}{2008}{\natexlab{b}}).

\bibitem[{\citenamefont{Ren et~al.}(2008)\citenamefont{Ren, Yang, Lu, Yi, Shen,
  Li, Che, Dong, Sun, Zhou et~al.}}]{ren08epl}
\bibinfo{author}{\bibfnamefont{Z.~A.} \bibnamefont{Ren}},
  \bibinfo{author}{\bibfnamefont{J.}~\bibnamefont{Yang}},
  \bibinfo{author}{\bibfnamefont{W.}~\bibnamefont{Lu}},
  \bibinfo{author}{\bibfnamefont{W.}~\bibnamefont{Yi}},
  \bibinfo{author}{\bibfnamefont{X.~L.} \bibnamefont{Shen}},
  \bibinfo{author}{\bibfnamefont{Z.~C.} \bibnamefont{Li}},
  \bibinfo{author}{\bibfnamefont{G.~C.} \bibnamefont{Che}},
  \bibinfo{author}{\bibfnamefont{X.~L.} \bibnamefont{Dong}},
  \bibinfo{author}{\bibfnamefont{L.~L.} \bibnamefont{Sun}},
  \bibinfo{author}{\bibfnamefont{F.}~\bibnamefont{Zhou}}, \bibnamefont{et~al.},
  \bibinfo{journal}{Europhys.\ Lett.} \textbf{\bibinfo{volume}{83}},
  \bibinfo{pages}{57002} (\bibinfo{year}{2008}).

\bibitem[{\citenamefont{Rotter et~al.}(2008)\citenamefont{Rotter, Tegel, and
  Johrendt}}]{rotter08}
\bibinfo{author}{\bibfnamefont{M.}~\bibnamefont{Rotter}},
  \bibinfo{author}{\bibfnamefont{M.}~\bibnamefont{Tegel}}, \bibnamefont{and}
  \bibinfo{author}{\bibfnamefont{D.}~\bibnamefont{Johrendt}},
  \bibinfo{journal}{Phys.\ Rev.\ Lett.} \textbf{\bibinfo{volume}{101}},
  \bibinfo{pages}{107006} (\bibinfo{year}{2008}).

\bibitem[{\citenamefont{Sasmal et~al.}(2008)\citenamefont{Sasmal, Lv, Lorenz,
  Guloy, Chen, Xue, and Chu}}]{sasmal08arxiv}
\bibinfo{author}{\bibfnamefont{K.}~\bibnamefont{Sasmal}},
  \bibinfo{author}{\bibfnamefont{B.}~\bibnamefont{Lv}},
  \bibinfo{author}{\bibfnamefont{B.}~\bibnamefont{Lorenz}},
  \bibinfo{author}{\bibfnamefont{A.}~\bibnamefont{Guloy}},
  \bibinfo{author}{\bibfnamefont{F.}~\bibnamefont{Chen}},
  \bibinfo{author}{\bibfnamefont{Y.}~\bibnamefont{Xue}}, \bibnamefont{and}
  \bibinfo{author}{\bibfnamefont{C.~W.} \bibnamefont{Chu}},
  \bibinfo{journal}{Phys.\ Rev.\ Lett.} \textbf{\bibinfo{volume}{101}},
  \bibinfo{pages}{107007} (\bibinfo{year}{2008}).

\bibitem[{\citenamefont{Wang et~al.}(2008)\citenamefont{Wang, Liu, Lv, Gao,
  Yang, Yu, Li, and Jin}}]{wang08arxiv}
\bibinfo{author}{\bibfnamefont{X.~C.} \bibnamefont{Wang}},
  \bibinfo{author}{\bibfnamefont{Q.~Q.} \bibnamefont{Liu}},
  \bibinfo{author}{\bibfnamefont{Y.~X.} \bibnamefont{Lv}},
  \bibinfo{author}{\bibfnamefont{W.~B.} \bibnamefont{Gao}},
  \bibinfo{author}{\bibfnamefont{L.~X.} \bibnamefont{Yang}},
  \bibinfo{author}{\bibfnamefont{R.~C.} \bibnamefont{Yu}},
  \bibinfo{author}{\bibfnamefont{F.~Y.} \bibnamefont{Li}}, \bibnamefont{and}
  \bibinfo{author}{\bibfnamefont{C.~Q.} \bibnamefont{Jin}}
  (\bibinfo{year}{2008}), \eprint{arXiv:0806.4688}.

\bibitem[{\citenamefont{Pitcher et~al.}(2008)\citenamefont{Pitcher, Parker,
  Adamson, Herkelrath, Boothroyd, and Clarke}}]{pitcher08arxiv}
\bibinfo{author}{\bibfnamefont{M.~J.} \bibnamefont{Pitcher}},
  \bibinfo{author}{\bibfnamefont{D.~R.} \bibnamefont{Parker}},
  \bibinfo{author}{\bibfnamefont{P.}~\bibnamefont{Adamson}},
  \bibinfo{author}{\bibfnamefont{S.~J.~C.} \bibnamefont{Herkelrath}},
  \bibinfo{author}{\bibfnamefont{A.~T.} \bibnamefont{Boothroyd}},
  \bibnamefont{and} \bibinfo{author}{\bibfnamefont{S.~J.} \bibnamefont{Clarke}}
  (\bibinfo{year}{2008}), \bibinfo{note}{(Chem. Commun. (2008) In Press)},
  \eprint{arXiv:0807.2228}.

\bibitem[{\citenamefont{Tapp et~al.}(2008)\citenamefont{Tapp, Tang, Lv, Sasmal,
  Lorenz, Chu, and Guloy}}]{tapp08prb}
\bibinfo{author}{\bibfnamefont{J.}~\bibnamefont{Tapp}},
  \bibinfo{author}{\bibfnamefont{Z.}~\bibnamefont{Tang}},
  \bibinfo{author}{\bibfnamefont{B.}~\bibnamefont{Lv}},
  \bibinfo{author}{\bibfnamefont{K.}~\bibnamefont{Sasmal}},
  \bibinfo{author}{\bibfnamefont{B.}~\bibnamefont{Lorenz}},
  \bibinfo{author}{\bibfnamefont{P.~C.~W.} \bibnamefont{Chu}},
  \bibnamefont{and} \bibinfo{author}{\bibfnamefont{A.~M.} \bibnamefont{Guloy}},
  \bibinfo{journal}{Phys.\ Rev.\ B} \textbf{\bibinfo{volume}{78}},
  \bibinfo{pages}{060505} (\bibinfo{year}{2008}).

\bibitem[{\citenamefont{Nowik and Felner}(2008)}]{nowik08arxiv}
\bibinfo{author}{\bibfnamefont{I.}~\bibnamefont{Nowik}} \bibnamefont{and}
  \bibinfo{author}{\bibfnamefont{I.}~\bibnamefont{Felner}}
  (\bibinfo{year}{2008}), \eprint{arXiv:0806.4078}.

\bibitem[{\citenamefont{Sidorenko et~al.}(2008)\citenamefont{Sidorenko, Renzi,
  Martinelli, and Palenzona}}]{sidorenko08arxiv}
\bibinfo{author}{\bibfnamefont{A.~A.} \bibnamefont{Sidorenko}},
  \bibinfo{author}{\bibfnamefont{R.~D.} \bibnamefont{Renzi}},
  \bibinfo{author}{\bibfnamefont{A.}~\bibnamefont{Martinelli}},
  \bibnamefont{and} \bibinfo{author}{\bibfnamefont{A.}~\bibnamefont{Palenzona}}
  (\bibinfo{year}{2008}), \eprint{arXiv:0807.0769}.

\bibitem[{\citenamefont{Yuzuri et~al.}(1980)\citenamefont{Yuzuri, Tahara, and
  Nakamura}}]{yuzuri80}
\bibinfo{author}{\bibfnamefont{M.}~\bibnamefont{Yuzuri}},
  \bibinfo{author}{\bibfnamefont{R.}~\bibnamefont{Tahara}}, \bibnamefont{and}
  \bibinfo{author}{\bibfnamefont{Y.}~\bibnamefont{Nakamura}},
  \bibinfo{journal}{J.\ Phys.\ Soc.\ Jpn.} \textbf{\bibinfo{volume}{48}},
  \bibinfo{pages}{1937} (\bibinfo{year}{1980}).

\bibitem[{\citenamefont{Blundell}(1999)}]{blundell99}
\bibinfo{author}{\bibfnamefont{S.~J.} \bibnamefont{Blundell}},
  \bibinfo{journal}{Contemp.\ Phys.} \textbf{\bibinfo{volume}{40}},
  \bibinfo{pages}{175} (\bibinfo{year}{1999}).

\bibitem[{\citenamefont{Sonier et~al.}(2000)\citenamefont{Sonier, Brewer, and
  Kiefl}}]{sonier00}
\bibinfo{author}{\bibfnamefont{J.~E.} \bibnamefont{Sonier}},
  \bibinfo{author}{\bibfnamefont{J.~H.} \bibnamefont{Brewer}},
  \bibnamefont{and} \bibinfo{author}{\bibfnamefont{R.~F.} \bibnamefont{Kiefl}},
  \bibinfo{journal}{Rev.\ Mod.\ Phys.} \textbf{\bibinfo{volume}{72}},
  \bibinfo{pages}{769} (\bibinfo{year}{2000}).

\bibitem[{\citenamefont{Luetkens
  et~al.}(2008{\natexlab{a}})\citenamefont{Luetkens, Klauss, Khasanov, Amato,
  Klingeler, Hellmann, Leps, Kondrat, Hess, Kohler et~al.}}]{luetkens08prl}
\bibinfo{author}{\bibfnamefont{H.}~\bibnamefont{Luetkens}},
  \bibinfo{author}{\bibfnamefont{H.~H.} \bibnamefont{Klauss}},
  \bibinfo{author}{\bibfnamefont{R.}~\bibnamefont{Khasanov}},
  \bibinfo{author}{\bibfnamefont{A.}~\bibnamefont{Amato}},
  \bibinfo{author}{\bibfnamefont{R.}~\bibnamefont{Klingeler}},
  \bibinfo{author}{\bibfnamefont{I.}~\bibnamefont{Hellmann}},
  \bibinfo{author}{\bibfnamefont{N.}~\bibnamefont{Leps}},
  \bibinfo{author}{\bibfnamefont{A.}~\bibnamefont{Kondrat}},
  \bibinfo{author}{\bibfnamefont{C.}~\bibnamefont{Hess}},
  \bibinfo{author}{\bibfnamefont{A.}~\bibnamefont{Kohler}},
  \bibnamefont{et~al.}, \bibinfo{journal}{Phys.\ Rev.\ Lett.}
  \textbf{\bibinfo{volume}{101}}, \bibinfo{pages}{097009}
  (\bibinfo{year}{2008}{\natexlab{a}}).

\bibitem[{\citenamefont{Drew et~al.}(2008{\natexlab{a}})\citenamefont{Drew,
  Pratt, Lancaster, Blundell, Baker, Liu, Wu, Chen, Watanabe, Malik
  et~al.}}]{drew08prl}
\bibinfo{author}{\bibfnamefont{A.~J.} \bibnamefont{Drew}},
  \bibinfo{author}{\bibfnamefont{F.~L.} \bibnamefont{Pratt}},
  \bibinfo{author}{\bibfnamefont{T.}~\bibnamefont{Lancaster}},
  \bibinfo{author}{\bibfnamefont{S.~J.} \bibnamefont{Blundell}},
  \bibinfo{author}{\bibfnamefont{P.~J.} \bibnamefont{Baker}},
  \bibinfo{author}{\bibfnamefont{R.~H.} \bibnamefont{Liu}},
  \bibinfo{author}{\bibfnamefont{G.}~\bibnamefont{Wu}},
  \bibinfo{author}{\bibfnamefont{X.~H.} \bibnamefont{Chen}},
  \bibinfo{author}{\bibfnamefont{I.}~\bibnamefont{Watanabe}},
  \bibinfo{author}{\bibfnamefont{V.~K.} \bibnamefont{Malik}},
  \bibnamefont{et~al.}, \bibinfo{journal}{Phys.\ Rev.\ Lett.}
  \textbf{\bibinfo{volume}{101}}, \bibinfo{pages}{097010}
  (\bibinfo{year}{2008}{\natexlab{a}}).

\bibitem[{\citenamefont{Carlo et~al.}(2008)\citenamefont{Carlo, Uemura, Goko,
  MacDougall, Rodriguez, Yu, Luke, Dai, Shannon, Miyasaka
  et~al.}}]{carlo08arxiv}
\bibinfo{author}{\bibfnamefont{J.~P.} \bibnamefont{Carlo}},
  \bibinfo{author}{\bibfnamefont{Y.~J.} \bibnamefont{Uemura}},
  \bibinfo{author}{\bibfnamefont{T.}~\bibnamefont{Goko}},
  \bibinfo{author}{\bibfnamefont{G.~J.} \bibnamefont{MacDougall}},
  \bibinfo{author}{\bibfnamefont{J.~A.} \bibnamefont{Rodriguez}},
  \bibinfo{author}{\bibfnamefont{W.}~\bibnamefont{Yu}},
  \bibinfo{author}{\bibfnamefont{G.~M.} \bibnamefont{Luke}},
  \bibinfo{author}{\bibfnamefont{P.}~\bibnamefont{Dai}},
  \bibinfo{author}{\bibfnamefont{N.}~\bibnamefont{Shannon}},
  \bibinfo{author}{\bibfnamefont{S.}~\bibnamefont{Miyasaka}},
  \bibnamefont{et~al.} (\bibinfo{year}{2008}), \eprint{arXiv:0805.2186}.

\bibitem[{\citenamefont{Luetkens
  et~al.}(2008{\natexlab{b}})\citenamefont{Luetkens, Klauss, Kraken, Litterst,
  Dellmann, Klingeler, Hess, Khasanov, Amato, Baines et~al.}}]{luetkens08arxiv}
\bibinfo{author}{\bibfnamefont{H.}~\bibnamefont{Luetkens}},
  \bibinfo{author}{\bibfnamefont{H.-H.} \bibnamefont{Klauss}},
  \bibinfo{author}{\bibfnamefont{M.}~\bibnamefont{Kraken}},
  \bibinfo{author}{\bibfnamefont{F.~J.} \bibnamefont{Litterst}},
  \bibinfo{author}{\bibfnamefont{T.}~\bibnamefont{Dellmann}},
  \bibinfo{author}{\bibfnamefont{R.}~\bibnamefont{Klingeler}},
  \bibinfo{author}{\bibfnamefont{C.}~\bibnamefont{Hess}},
  \bibinfo{author}{\bibfnamefont{R.}~\bibnamefont{Khasanov}},
  \bibinfo{author}{\bibfnamefont{A.}~\bibnamefont{Amato}},
  \bibinfo{author}{\bibfnamefont{C.}~\bibnamefont{Baines}},
  \bibnamefont{et~al.} (\bibinfo{year}{2008}{\natexlab{b}}),
  \eprint{arXiv:0806.3533}.

\bibitem[{\citenamefont{Khasanov et~al.}(2008)\citenamefont{Khasanov, Luetkens,
  Amato, Klauss, Ren, Yang, Lu, and Zhao}}]{khasanov08arxiv}
\bibinfo{author}{\bibfnamefont{R.}~\bibnamefont{Khasanov}},
  \bibinfo{author}{\bibfnamefont{H.}~\bibnamefont{Luetkens}},
  \bibinfo{author}{\bibfnamefont{A.}~\bibnamefont{Amato}},
  \bibinfo{author}{\bibfnamefont{H.-H.} \bibnamefont{Klauss}},
  \bibinfo{author}{\bibfnamefont{Z.-A.} \bibnamefont{Ren}},
  \bibinfo{author}{\bibfnamefont{J.}~\bibnamefont{Yang}},
  \bibinfo{author}{\bibfnamefont{W.}~\bibnamefont{Lu}}, \bibnamefont{and}
  \bibinfo{author}{\bibfnamefont{Z.-X.} \bibnamefont{Zhao}}
  (\bibinfo{year}{2008}), \eprint{arXiv:0805.1923}.

\bibitem[{\citenamefont{Takeshita et~al.}(2008)\citenamefont{Takeshita, Kadono,
  Hiraishi, Miyazaki, Koda, Kamihara, and Hosono}}]{takeshita08arxiv}
\bibinfo{author}{\bibfnamefont{S.}~\bibnamefont{Takeshita}},
  \bibinfo{author}{\bibfnamefont{R.}~\bibnamefont{Kadono}},
  \bibinfo{author}{\bibfnamefont{M.}~\bibnamefont{Hiraishi}},
  \bibinfo{author}{\bibfnamefont{M.}~\bibnamefont{Miyazaki}},
  \bibinfo{author}{\bibfnamefont{A.}~\bibnamefont{Koda}},
  \bibinfo{author}{\bibfnamefont{Y.}~\bibnamefont{Kamihara}}, \bibnamefont{and}
  \bibinfo{author}{\bibfnamefont{H.}~\bibnamefont{Hosono}}
  (\bibinfo{year}{2008}), \eprint{arXiv:0806.4798}.

\bibitem[{\citenamefont{Drew et~al.}(2008{\natexlab{b}})\citenamefont{Drew,
  Niedermayer, Baker, Pratt, Blundell, Lancaster, Liu, Wu, Chen, Watanabe
  et~al.}}]{drew08arxiv}
\bibinfo{author}{\bibfnamefont{A.~J.} \bibnamefont{Drew}},
  \bibinfo{author}{\bibfnamefont{C.}~\bibnamefont{Niedermayer}},
  \bibinfo{author}{\bibfnamefont{P.~J.} \bibnamefont{Baker}},
  \bibinfo{author}{\bibfnamefont{F.~L.} \bibnamefont{Pratt}},
  \bibinfo{author}{\bibfnamefont{S.~J.} \bibnamefont{Blundell}},
  \bibinfo{author}{\bibfnamefont{T.}~\bibnamefont{Lancaster}},
  \bibinfo{author}{\bibfnamefont{R.~H.} \bibnamefont{Liu}},
  \bibinfo{author}{\bibfnamefont{G.}~\bibnamefont{Wu}},
  \bibinfo{author}{\bibfnamefont{X.~H.} \bibnamefont{Chen}},
  \bibinfo{author}{\bibfnamefont{I.}~\bibnamefont{Watanabe}},
  \bibnamefont{et~al.} (\bibinfo{year}{2008}{\natexlab{b}}),
  \eprint{arXiv:0807.4876}.

\bibitem[{\citenamefont{Aczel et~al.}(2008)\citenamefont{Aczel,
  Baggio-Saitovitch, Budko, Canfield, Carlo, Chen, Dai, Goko, Hu, Luke
  et~al.}}]{aczel08arxiv}
\bibinfo{author}{\bibfnamefont{A.~A.} \bibnamefont{Aczel}},
  \bibinfo{author}{\bibfnamefont{E.}~\bibnamefont{Baggio-Saitovitch}},
  \bibinfo{author}{\bibfnamefont{S.~L.} \bibnamefont{Budko}},
  \bibinfo{author}{\bibfnamefont{P.~C.} \bibnamefont{Canfield}},
  \bibinfo{author}{\bibfnamefont{J.~P.} \bibnamefont{Carlo}},
  \bibinfo{author}{\bibfnamefont{G.~F.} \bibnamefont{Chen}},
  \bibinfo{author}{\bibfnamefont{P.}~\bibnamefont{Dai}},
  \bibinfo{author}{\bibfnamefont{T.}~\bibnamefont{Goko}},
  \bibinfo{author}{\bibfnamefont{W.~Z.} \bibnamefont{Hu}},
  \bibinfo{author}{\bibfnamefont{G.~M.} \bibnamefont{Luke}},
  \bibnamefont{et~al.} (\bibinfo{year}{2008}), \eprint{arXiv:0807.1044}.

\bibitem[{\citenamefont{Goko et~al.}(2008)\citenamefont{Goko, Aczel,
  Baggio-Saitovitch, Bud'ko, Canfield, Carlo, Chen, Dai, Hamann, Hu
  et~al.}}]{goko08arxiv}
\bibinfo{author}{\bibfnamefont{T.}~\bibnamefont{Goko}},
  \bibinfo{author}{\bibfnamefont{A.~A.} \bibnamefont{Aczel}},
  \bibinfo{author}{\bibfnamefont{E.}~\bibnamefont{Baggio-Saitovitch}},
  \bibinfo{author}{\bibfnamefont{S.~L.} \bibnamefont{Bud'ko}},
  \bibinfo{author}{\bibfnamefont{P.~C.} \bibnamefont{Canfield}},
  \bibinfo{author}{\bibfnamefont{J.~P.} \bibnamefont{Carlo}},
  \bibinfo{author}{\bibfnamefont{G.~F.} \bibnamefont{Chen}},
  \bibinfo{author}{\bibfnamefont{P.}~\bibnamefont{Dai}},
  \bibinfo{author}{\bibfnamefont{A.~C.} \bibnamefont{Hamann}},
  \bibinfo{author}{\bibfnamefont{W.~Z.} \bibnamefont{Hu}}, \bibnamefont{et~al.}
  (\bibinfo{year}{2008}), \eprint{arXiv:0808.1425}.

\bibitem[{\citenamefont{Klauss et~al.}(2008)\citenamefont{Klauss, Luetkens,
  Klingeler, Hess, Litterst, Kraken, Korshunov, Eremin, Drechsler, Khasanov
  et~al.}}]{klauss08prl}
\bibinfo{author}{\bibfnamefont{H.-H.} \bibnamefont{Klauss}},
  \bibinfo{author}{\bibfnamefont{H.}~\bibnamefont{Luetkens}},
  \bibinfo{author}{\bibfnamefont{R.}~\bibnamefont{Klingeler}},
  \bibinfo{author}{\bibfnamefont{C.}~\bibnamefont{Hess}},
  \bibinfo{author}{\bibfnamefont{F.~J.} \bibnamefont{Litterst}},
  \bibinfo{author}{\bibfnamefont{M.}~\bibnamefont{Kraken}},
  \bibinfo{author}{\bibfnamefont{M.~M.} \bibnamefont{Korshunov}},
  \bibinfo{author}{\bibfnamefont{I.}~\bibnamefont{Eremin}},
  \bibinfo{author}{\bibfnamefont{S.-L.} \bibnamefont{Drechsler}},
  \bibinfo{author}{\bibfnamefont{R.}~\bibnamefont{Khasanov}},
  \bibnamefont{et~al.}, \bibinfo{journal}{Phys.\ Rev.\ Lett.}
  \textbf{\bibinfo{volume}{101}}, \bibinfo{pages}{077005}
  (\bibinfo{year}{2008}).

\bibitem[{foo({\natexlab{a}})}]{footnote1}
\bibinfo{note}{The possibility of finding \chem{Fe_{2}As}, which has $T_{\rm N}
  = 353$~K has also been discussed~\cite{nowik08arxiv}, but since no magnetic
  signals above $160$K have been reported we have not investigated this
  compound.}

\bibitem[{\citenamefont{Selte et~al.}(1972)\citenamefont{Selte, Kjekshus, and
  Andresen}}]{selte72}
\bibinfo{author}{\bibfnamefont{K.}~\bibnamefont{Selte}},
  \bibinfo{author}{\bibfnamefont{A.}~\bibnamefont{Kjekshus}}, \bibnamefont{and}
  \bibinfo{author}{\bibfnamefont{A.~F.} \bibnamefont{Andresen}},
  \bibinfo{journal}{Acta.\ Chem.\ Scand.} \textbf{\bibinfo{volume}{26}},
  \bibinfo{pages}{3101} (\bibinfo{year}{1972}).

\bibitem[{\citenamefont{Lyman and Prewitt}(1984)}]{lyman84}
\bibinfo{author}{\bibfnamefont{P.~S.} \bibnamefont{Lyman}} \bibnamefont{and}
  \bibinfo{author}{\bibfnamefont{C.~T.} \bibnamefont{Prewitt}},
  \bibinfo{journal}{Acta Cryst. B} \textbf{\bibinfo{volume}{40}},
  \bibinfo{pages}{14} (\bibinfo{year}{1984}).

\bibitem[{\citenamefont{Fan et~al.}(1972)\citenamefont{Fan, Rosenthal,
  McKinzie, and Wold}}]{fan72}
\bibinfo{author}{\bibfnamefont{A.~K.~L.} \bibnamefont{Fan}},
  \bibinfo{author}{\bibfnamefont{G.~H.} \bibnamefont{Rosenthal}},
  \bibinfo{author}{\bibfnamefont{H.~L.} \bibnamefont{McKinzie}},
  \bibnamefont{and} \bibinfo{author}{\bibfnamefont{A.}~\bibnamefont{Wold}},
  \bibinfo{journal}{J. Solid. State Chem.} \textbf{\bibinfo{volume}{5}},
  \bibinfo{pages}{136} (\bibinfo{year}{1972}).

\bibitem[{foo({\natexlab{b}})}]{footnote2}
\bibinfo{note}{We have confirmed this by carrying out test measurements of
  \chem{FeAs} at the ISIS Pulsed Muon Facility, UK.}

\end{thebibliography}

\end{document}